\documentclass[aps,prl,twocolumn,nofootinbib,superscriptaddress,amsmath,amssymb,floatfix]{revtex4}

\usepackage{graphicx}
\usepackage{dcolumn}
\usepackage{bm,color}
\usepackage{amssymb}
\usepackage[colorlinks=true,citecolor=red,urlcolor=black,linkcolor=blue]{hyperref}

\allowdisplaybreaks

\begin{document}

\title{The complete theory of Maxwell and Proca fields}
\author{Ver\'onica Errasti D\'iez}
\email{ veroerdi@mppmu.mpg.de}
\author{Brage Gording}
\email{ brageg@mppmu.mpg.de}
\author{Julio A. M\'{e}ndez-Zavaleta}
\email{julioamz@mpp.mpg.de}
\author{Angnis Schmidt-May}
\email{angnissm@mpp.mpg.de}
\affiliation{Max-Planck-Institut f\"ur Physik (Werner-Heisenberg-Institut),
F\"ohringer Ring 6, 80805 Munich, Germany}

%%%%%%%%%%%%%%%%%%%%%%%%%%%%%%%%%%%%%%%%%%%%%%%%%%%%%%%%%
%ABSTRACT. Words: 169 (Must be under 600)

\begin{abstract}
We present the most general ghost-free classical Lagrangian containing first-order derivatives and describing interacting real Abelian spin-one fields on Minkowski spacetime.
We study both massive Proca and massless Maxwell fields and allow for a non-linear realization of mass, in the form of derivative self-interactions.
Within this context, our construction notoriously extends the existing literature,
which is limited to the case of a single Proca field and to multiple interacting Proca fields in the presence of a global rotational symmetry.
In the limit of a single Proca field, we reproduce the known healthy interaction terms.
We provide the necessary and sufficient conditions to ensure ghost-freedom in any multi-field setup.
We observe that, in general, the said conditions are not satisfied by the rotationally symmetric multi-Proca interactions suggested so far,
which implies that they propagate ghosts.
Our theory admits a plethora of applications in a wide range of subjects.
For illustrative purposes, we provide concrete proposals in holographic condensed matter and black hole physics.
\end{abstract}

\maketitle

%%%%%%%%%%%%%%%%%%%%%%%%%%%%%%%%%%%%%%%%%%%%%%%%%%%%%%%%%
%INTRODUCTION. Total words: 864

\section{Introduction}
\label{intro}

%%%%%%%%%%%%%%%%%%%%%%%%%%%%%%%%%%%%%%%%%%%%%%%%%%%%%%%%%
%INTRODUCTION, 1st paragraph. Words: 145

The investigation of four-dimensional consistent theories of fields of any spin has been a subject of great interest over many decades
and continues to be zealously pursued.
This research line started in~\cite{FPau} where, based on Lorentz invariance and positivity of energy,
the Fierz-Pauli equations describing the dynamics of free massive fields of arbitrary spin were obtained.
It was not until much later that a proper Lagrangian formulation of these equations was established~\cite{SHagen}.
The case of free massless fields was formalized in~\cite{Frons}, where both the Fronsdal equations of motion and the corresponding action principle were derived.
In spite of noteworthy and ongoing progress, the extension of these setups to the case of interacting fields remains evasive.
One of the reasons is that, generically, the addition of interaction terms to a Lagrangian introduces further unphysical degrees of freedom.
Throughout this paper, we refer to these as ghosts.

%%%%%%%%%%%%%%%%%%%%%%%%%%%%%%%%%%%%%%%%%%%%%%%%%%%%%%%%%
%INTRODUCTION, 2nd paragraph. Words: 112

Generalizations that include interactions are particularly challenging when the spin of at least one field is greater than two
---which collectively are known as higher-spin theories.
Indeed, consistent classical theories for interacting fields exist only for spin $s\leq2$.
This is because, in the presence of one or more such fields, the spectrum of the corresponding theory is necessarily infinite,
i.e.~it contains fields of all spins.
Although generic non-linear equations of motion capturing higher-spin interactions do exist~\cite{Vasi},
so far no complete action has been constructed to describe their dynamics.
Nonetheless, effective field theories for a single massive higher-spin particle are possible, e.g.~\cite{Bellazzini:2019bzh}.
(For enlightening reviews on higher-spin theories, see~\cite{RevSp} and references therein.)

%%%%%%%%%%%%%%%%%%%%%%%%%%%%%%%%%%%%%%%%%%%%%%%%%%%%%%%%%
%INTRODUCTION, 3rd paragraph. Words: 119

The case of spin-two fields has met with more success owing to the fact that, when restricting to Einstein-Hilbert kinetic terms,
the corresponding Lagrangian admits a finite closed form.
The linear theory of a single massless spin-two field in a flat background propagating the correct number of physical degrees of freedom~\cite{FPau,SHagen}
admits a most celebrated non-linear completion: General Relativity.
Its massive counterpart was first proposed in~\cite{deRham:2010kj} and shortly after proven to be ghost-free~\cite{Hassan:2011hr}.
Powerful theorems forbid the existence of consistent theories of interacting massless spin-two fields~\cite{Boulanger:2000rq}.
Consequently, the next advancement in this context involved formulating theories of multiple interacting spin-two fields,
where at most only one such field is allowed to be massless~\cite{Hinterbichler:2012cn}.
These theories are still being developed~\cite{Hassan:2018mcw}.

%%%%%%%%%%%%%%%%%%%%%%%%%%%%%%%%%%%%%%%%%%%%%%%%%%%%%%%%%
%INTRODUCTION, 4th paragraph. Words: 78

Turning attention to spin-zero fields, one observes that (multi-)Galileon theories have been extensively studied, see e.g.~\cite{Galileons}.
In brief, these are consistent theories of interacting scalar fields on Minkowski spacetime
whose Lagrangian is invariant under a shift symmetry and leads to second-order equations of motion.
Galileons have also been studied over curved spacetimes~\cite{curvGal}, where they match the well-known Horndeski theories~\cite{Horn}
and have inspired the more modern beyond-Horndeski theories~\cite{Gleyzes:2014dya}.
The latter include interactions that lead to higher-order equations of motion.

%%%%%%%%%%%%%%%%%%%%%%%%%%%%%%%%%%%%%%%%%%%%%%%%%%%%%%%%%
%INTRODUCTION, 5th paragraph. Words: 193

Compared to their spin-two and spin-zero analogues, healthy classical theories of spin-one fields are still poorly understood, even when one focuses solely on real Abelian fields $X_\mu$.
In this spin-one case, we focus on first-order theories only.
By first-order we mean that the Lagrangian contains exclusively first-order derivatives and powers of them but no higher derivatives,
up to integration by parts:
$\mathcal{L}=\mathcal{L}(X_\mu,\partial_\mu X_\nu)$.
What broadly is referred to as non-linear electrodynamics~\cite{Pleb} accounts for the completion of Maxwell's theory
of a single massless Abelian spin-one field.
This includes the famous Born-Infeld Lagrangian~\cite{BI}, but also more recent proposals
such as exponential~\cite{Hendi} and logarithmic~\cite{Neto} electrodynamics.
On the massive side, the Proca action~\cite{Proca} started to be supplemented with derivative self-interaction terms not long ago~\cite{Tasinato:2014eka}.
Little is known when it comes to multiple fields.
Purely massless interactions are possible at the spin-one level because the no-go theorems in~\cite{Boulanger:2000rq} no longer apply,
but we were not able to find any reference dedicated to their study.
Purely massive interactions have only been studied in~\cite{Jimenez:2016upj}, to our knowledge.
However, a very stringent global symmetry on the field space was there imposed, yielding a theory much more restricted than the general case.

%%%%%%%%%%%%%%%%%%%%%%%%%%%%%%%%%%%%%%%%%%%%%%%%%%%%%%%%%
%INTRODUCTION, 6th paragraph. Words: 139

{\it Summary of results.}
In this work, we fill the void and introduce the complete consistent classical theory of interacting real Abelian spin-one fields on Minkowski spacetime.
We allow for both massless and massive fields
and introduce the most general first-order Lagrangian that contains (self-)interaction terms among all fields.
Notably, when multiple massive fields are considered,
the interaction terms we put forward must satisfy certain differential relations in order to ensure ghost-freedom.
These relations guarantee the existence of the secondary second class constraint inherent to every massive field.
We point out that the subset of purely massive interactions in~\cite{Jimenez:2016upj} does not always fulfil the said relations and so it generically propagates ghosts.
The Lagrangian we propose contains the most general theory of a single Proca field as a subcase.
In this limit, we replicate previous proposals in the literature.

%%%%%%%%%%%%%%%%%%%%%%%%%%%%%%%%%%%%%%%%%%%%%%%%%%%%%%%%%
%INTRODUCTION, 7th paragraph. Words: 78

{\it Conventions.}
We work on Minkowski spacetime.
We choose Cartesian coordinates with the mostly positive signature $\eta_{\mu\nu}=diag(-1,1,1,1)$.
We use the convention that Greek alphabets $(\mu,\nu,\dots)$ denote spacetime indices and are raised/lowered by $\eta_{\mu\nu}$ and its inverse $\eta^{\mu\nu}$.
The soon to be introduced labels $(A_i,\alpha,\overline{\alpha})$ enumerate fields.
Raising and lowering these labels is trivial, since we do not impose any symmetry on the field space.
As usual, sum over repeated indices and labels should be understood at all times.

%%%%%%%%%%%%%%%%%%%%%%%%%%%%%%%%%%%%%%%%%%%%%%%%%%%%%%%%%
%MAIN FEATURES. Total words: 1269

\section{Main features}
\label{sectmaxpr}

%%%%%%%%%%%%%%%%%%%%%%%%%%%%%%%%%%%%%%%%%%%%%%%%%%%%%%%%%
%MAIN FEATURES, 1st paragraph. Words: 103

Real Abelian spin-one fields can be either massless or massive.
These are widely known as Maxwell and Proca fields, respectively.
At a rigorous algebraic level, a Maxwell field is characterized by two first class constraints
while a Proca field is associated with two second class constraints: one primary and one secondary.
Interestingly, the latter algebraic definition is oblivious to the precise realization of the second class constraints:
as an explicit mass term in the Lagrangian~\cite{Proca} and/or as derivative self-interaction terms~\cite{Tasinato:2014eka}.
The latter case is also known as generalized Proca or vector-Galileon.
Subsequently, we adopt the algebraic understanding for both Maxwell and Proca fields.

%%%%%%%%%%%%%%%%%%%%%%%%%%%%%%%%%%%%%%%%%%%%%%%%%%%%%%%%%
%MAIN FEATURES, 2nd paragraph. Words: 78

We consider $N$ number of Maxwell fields and $M$ number of Proca fields on Minkowski spacetime $\mathbb{R}\times\mathbb{R}^3$.
We allow all $N+M$ fields to self-interact as well as couple to each other in the most general manner that avoids ghost-like degrees of freedom.
We restrict attention to first-order (self-)interactions.
Of course, these lead to at most second-order equations of motion, a feature which ensures that the Ostrogradsky instability~\cite{Ostr} is avoided.
We refer to our result as the Maxwell-Proca theory.

%%%%%%%%%%%%%%%%%%%%%%%%%%%%%%%%%%%%%%%%%%%%%%%%%%%%%%%%%
%MAIN FEATURES, 3rd paragraph. Words: 198

Explicitly, we take all Maxwell $\{A_\mu\}$ and Proca $\{B_\mu\}$ fields to be real.
Their field strengths are
\begin{align}
A_{\mu\nu}^{(\overline{\alpha})}=\partial_\mu A_\nu^{(\overline{\alpha})}-\partial_\nu A_\mu^{(\overline{\alpha})}, \quad
B_{\mu\nu}^{(\alpha)}=\partial_\mu B_\nu^{(\alpha)}-\partial_\nu B_\mu^{(\alpha)}, \label{maxpr}
\end{align}
for all $\overline{\alpha}=1,\ldots,N$ and $\alpha=1,\dots,M$.
We write the action for the Maxwell-Proca theory as
\begin{align}
S=\int_{\mathbb{R}\times \mathbb{R}^3}\textrm{d}^{4}x\,\, \mathcal{L}_{\textrm{MP}}, \quad \mathcal{L}_{\textrm{MP}}=\mathcal{L}_{\textrm{kin}}+\mathcal{L}_{\textrm{int}}, \label{fullLag}
\end{align}
where the kinetic piece of the Lagrangian is canonically normalized,
\begin{align}
\mathcal{L}_{\textrm{kin}}=-\frac{1}{4}A_{\mu\nu}^{(\overline{\alpha})}A^{\mu\nu}_{(\overline{\alpha})}-\frac{1}{4}B_{\mu\nu}^{(\alpha)}
B^{\mu\nu}_{(\alpha)}, \label{kin}
\end{align}
and we split the interaction piece into three parts: interactions among the Maxwell fields $\mathcal{L}^{(AA)}$,
between the Maxwell and Proca fields $\mathcal{L}^{(AB)}$ and
among the Proca fields $\mathcal{L}^{(BB)}$;
\begin{align}
\mathcal{L}_{\textrm{int}}=\mathcal{L}^{(AA)}+\mathcal{L}^{(AB)}+\mathcal{L}^{(BB)}. \label{intlag}
\end{align}
The Maxwell-Proca action depends on the Maxwell fields exclusively through their field strengths and so it has a manifest $U(1)^N$ gauge symmetry.
The dependence on the Proca fields is less restricted and consists of non-linear functions of the fields themselves,
powers of their first-order derivatives, field strengths and combinations thereof.
Consequently, no gauge symmetry is associated with the Proca fields.
Rather, it is explicitly broken.

%%%%%%%%%%%%%%%%%%%%%%%%%%%%%%%%%%%%%%%%%%%%%%%%%%%%%%%%%
%MAIN FEATURES, 4th paragraph. Words: 294

In more detail, $\mathcal{L}^{(AA)}$ encodes the (straightforward) extension to multiple fields of the already mentioned non-linear electrodynamics~\cite{Pleb,BI,Hendi,Neto}.
We can succinctly write these interactions as an arbitrary smooth real function of the Maxwell field strengths:
\begin{align}
\mathcal{L}^{(AA)}=f\big(A_{\mu\nu}^{(\overline{\alpha})}\big). \label{LAA}
\end{align}
On the other hand, $\mathcal{L}^{(AB)}$ and $\mathcal{L}^{(BB)}$ present a rich novel structure.
This is best understood by means of the subdivision
\begin{align}
\mathcal{L}^{(AB)}=\sum_{n=0}^4\mathcal{L}^{(AB)}_{(n)}, \quad \mathcal{L}^{(BB)}=\sum_{n=0}^4\mathcal{L}^{(BB)}_{(n)}.
\label{sumn}
\end{align}
Let $X$ stand for either a Maxwell or a Proca field:
\begin{align}
\{X^{(A_i)}\}=\{A^{(\overline{\alpha})}\}\cup\{B^{(\alpha)}\}, \label{sets}
\end{align}
where the field labels $A_i$ range from 1 to $N+M$.
Then, $\mathcal{L}_{(0)}^{(XB)}$ in (\ref{sumn}) denotes all interaction terms
that depend on derivatives of the fields exclusively through their field strengths:
\begin{align}
\mathcal{L}^{(AB)}_{(0)}=g\big(B_{\mu}^{(\alpha)},A_{\mu\nu}^{(\overline{\alpha})},B_{\mu\nu}^{(\alpha)}\big), \quad
\mathcal{L}^{(BB)}_{(0)}=h\big(B_{\mu}^{(\alpha)},B_{\mu\nu}^{(\alpha)}\big), \label{L0gh}
\end{align}
with $(g,h)$ arbitrary smooth real functions.
In $\mathcal{L}_{(n>0)}^{(XB)}$, the integer $n$ counts the number of derivatives.
These interactions are of the form
\begin{align}
\mathcal{L}_{(n>0)}^{(XX)}\!=\mathcal{T}^{\nu_1\dots\nu_n\rho_1\dots\rho_n}_{A_1\dots A_n}
\partial_{\nu_1}\! X^{(A_1)}_{\rho_1}\dots\partial_{\nu_n}\! X^{(A_n)}_{\rho_n}. \label{Lng0int}
\end{align}
Here, $\mathcal{T}^{\nu_1\dots\nu_n\rho_1\dots\rho_n}_{A_1\dots A_n}$
---$\mathcal{T}$ for short--- is a smooth real object constructed out of
the spacetime metric $\eta_{\mu\nu}$, the four-dimensional Levi-Civita tensor $\epsilon_{\mu\nu\rho\sigma}$ and the Proca fields $B_\mu^{(\alpha)}$.
Notably, $\mathcal{T}$ is such that the derivatives $\partial_{\nu_i}\! X^{(A_i)}_{\rho_i}$ are antisymmetrized so that,
when a time derivative acts on a temporal component $\partial_0X_0^{(A_i)}$, no other time derivatives appear.
This is the necessary and sufficient condition to avoid the propagation of ghosts at the primary level of the constraint algebra.
As already pointed out, when a Maxwell field appears in (\ref{Lng0int}), it does so through its field strength.

%%%%%%%%%%%%%%%%%%%%%%%%%%%%%%%%%%%%%%%%%%%%%%%%%%%%%%%%%
%MAIN FEATURES, 5th paragraph. Words: 83

A couple of remarks are due.
First, if one restricts attention to four-dimensional first-order theories, the two series in (\ref{sumn}) stop at $n=4$~\cite{footnmax}.
Second, note that $\mathcal{L}^{(AB)}_{(1)}\subseteq \mathcal{L}^{(AB)}_{(0)}$~\cite{footLAB1}.
In the same spirit, the generic form of the $\mathcal{L}_{(n>0)}^{(XX)}$ interactions in (\ref{Lng0int}) contains terms that belong in $\mathcal{L}_{(0)}^{(XX)}$ by definition.
Although no term is missed, there is a clear over-counting in this schematic presentation of the proposed interactions.
Such redundancies are further discussed and altogether removed by a close inspection of the $\mathcal{T}$'s elsewhere~\cite{PRD}.

%%%%%%%%%%%%%%%%%%%%%%%%%%%%%%%%%%%%%%%%%%%%%%%%%%%%%%%%%
%MAIN FEATURES, 6th paragraph. Words: 174

Crucially, all $\mathcal{T}$'s in (\ref{Lng0int})
must satisfy certain differential relations so that the Maxwell-Proca theory does not propagate ghosts.
The said relations are the necessary and sufficient condition for ensuring ghost-freedom at the secondary level of the constraint algebra.
To illustrate this point, consider two Proca fields interacting with each other via
\begin{align}
\mathcal{L}_{\textrm{int}}=f_{(1)}\partial\cdot B^{(1)}+f_{(2)}\partial\cdot B^{(2)}\in \mathcal{L}_{(1)}^{(BB)}, \label{exint}
\end{align}
with $f_{(1)}$ and $f_{(2)}$ smooth real functions of $B^{(1)}$ and $B^{(2)}$.
The above interactions propagate a ghost-like degree of freedom unless the relation
\begin{align}
\frac{\partial f_{(1)}}{\partial B^{(2)}_0}-\frac{\partial f_{(2)}}{\partial B^{(1)}_0}=0\label{excons}
\end{align}
holds true. Obviously, setting both $f_{(1)}$ and $f_{(2)}$ to constants satisfies (\ref{excons}), but this choice renders (\ref{exint}) irrelevant:
it reduces to a boundary term.
A more involved example of functions fulfilling (\ref{excons}) is given by
\begin{align}
\label{concrexf}
\begin{cases}
f_{(1)}=B^{(1)}\cdot B^{(2)}+\frac{1}{2}B^{(2)}\cdot B^{(2)}, \\
f_{(2)}=B^{(1)}\cdot B^{(2)}+\frac{1}{2}B^{(1)}\cdot B^{(1)}.
\end{cases}
\end{align}

%%%%%%%%%%%%%%%%%%%%%%%%%%%%%%%%%%%%%%%%%%%%%%%%%%%%%%%%%
%MAIN FEATURES, 7th paragraph. Words: 230

The secondary constraint enforcing relations such as (\ref{excons}) are an intrinsically multi-Proca result, arising when $M\geq2$.
For any number of Maxwell fields, they are trivially satisfied when a single Proca field is considered and also when Proca fields do not interact with each other.
At the constraint algebra level, they ensure that the secondary second class constraint inherent to every Proca field exists.
In full generality, they read
\begin{align}
\frac{\partial^2\mathcal{L}^{(XB)}_{(n)}}{\partial \dot{B}_0^{(\alpha)}\partial B_0^{(\beta)}}
-\frac{\partial^2\mathcal{L}^{(XB)}_{(n)}}{\partial \dot{B}_0^{(\beta)}\partial B_0^{(\alpha)}}=0 \qquad \forall n,\alpha,\beta, \label{secconst}
\end{align}
where once again $X$ spans all the Maxwell and Proca fields, as in (\ref{sets}).
To our mind, this is our strongest result, because all literature so far is implicitly limited to the study of the primary level of the constraint algebra.
As it turns out, with no thorough exploration of the secondary level,
it is simply impossible to guarantee ghost-freedom in any multi-field setup involving at least two Proca fields.
In particular, when $N=0$ and $M$ Proca fields interact in a globally symmetric manner, ghosts are generically propagated unless (\ref{secconst}) is enforced.
Therefore, the general interactions proposed in~\cite{Jimenez:2016upj} must be supplemented by (\ref{secconst})
before the resulting theory can be claimed healthy.
Specific interactions shown there are incompatible with these relations and hence propagate ghosts.
This is discussed and exemplified elsewhere~\cite{PRD}.
In all cases, once (\ref{secconst}) is satisfied, the constraint algebra automatically closes at the tertiary level.

%%%%%%%%%%%%%%%%%%%%%%%%%%%%%%%%%%%%%%%%%%%%%%%%%%%%%%%%%
%MAIN FEATURES, 8th paragraph. Words: 73

On the whole, the defining property of the Maxwell-Proca theory is its constraint algebra structure:
$N$ times that of a Maxwell field, plus $M$ times that of a Proca.
Namely, the infinitely many interaction terms (\ref{intlag}) we propose do not spoil the additivity and associativity of the constituent constraint algebras:
those of a single Maxwell and Proca field.
As a consequence, our theory propagates $2N+3M$ degrees of freedom and ghosts are unequivocally avoided.

%%%%%%%%%%%%%%%%%%%%%%%%%%%%%%%%%%%%%%%%%%%%%%%%%%%%%%%%%
%MAIN FEATURES, 9th paragraph. Words: 35

{\it The single Proca limit.}
When a single Proca field $B_{\mu}$ is considered ($N=0$ and $M=1$),
the Maxwell-Proca theory reduces to the Lagrangian reported in~\cite{Jimenez:2016isa,Allys:2016jaq,Heisenberg:2017mzp}.
Our construction thus reproduces the complete theory for a single Proca field~\cite{footsingle}.

%%%%%%%%%%%%%%%%%%%%%%%%%%%%%%%%%%%%%%%%%%%%%%%%%%%%%%%%%
%APPLICATIONS. Total words: 1172

\section{Concrete applications}
\label{holsect}

%%%%%%%%%%%%%%%%%%%%%%%%%%%%%%%%%%%%%%%%%%%%%%%%%%%%%%%%%
%APPLICATIONS, 1st paragraph. Words: 106
It is natural to regard the Maxwell-Proca theory here presented as a multi-field extension of classical electrodynamics
that includes non-linear (self-)interactions among the fields.
The usual Maxwell Lagrangian is given by the $N=1$ and $M=0$ limit of the Maxwell-Proca theory,
with no interaction terms at all: $\mathcal{L}_{\textrm{int}}=0$.
Since classical electrodynamics is one of the cornerstones of theoretical physics and since most physical phenomena are non-linear in nature,
it is not surprising that the Maxwell-Proca theory offers lush possibilities in diverse contexts.
In the following, we outline two definite applications that rely on the multi-field feature:
one in holographic condensed matter and one in black hole physics.

%%%%%%%%%%%%%%%%%%%%%%%%%%%%%%%%%%%%%%%%%%%%%%%%%%%%%%%%%
%APPLICATIONS, 2nd paragraph. Words: 135

{\it Coupling to gravity.}
Besides being an intrinsically interesting addition to the set of consistent classical field theories,
the Maxwell-Proca theory serves as a building block for the construction of more general healthy theories as well.
Indeed, efforts are made not only to derive ghost-free interaction terms among particles of the same spin,
but also between particles of different spins.
In this respect, the consistent coupling of the Maxwell-Proca theory to gravity stands out as a challenging yet fruitful scenario to investigate.
Challenging because such a coupling requires a rigorous Hamiltonian analysis to unambiguously avoid ghosts.
In fact, this analysis has already been initiated and in~\cite{Hull:2015uwa} the single Proca case was studied,
but without aiming for exhaustiveness.
Fruitful because, if successful ---even for simple subcases of the Maxwell-Proca theory---, the subsequent applications emerge.

%%%%%%%%%%%%%%%%%%%%%%%%%%%%%%%%%%%%%%%%%%%%%%%%%%%%%%%%%
%APPLICATIONS, 3rd paragraph. Words: 97

{\it Holographic condensed matter.}
The so-called AdS/CFT correspondence~\cite{Maldacena:1997re} establishes a duality between certain quantum systems and classical theories of gravity in one more dimension.
Since the correspondence relates the strongly coupled regime of one theory to the weakly coupled regime of its dual,
it is often employed as a tool in the resolution of otherwise untackable problems.
The basic idea is to address a question within strongly correlated quantum systems using perturbative techniques in gravity.
Accordingly, holographic condensed matter refers to the study of microscopic properties of (often idealized) materials in terms of the dual gravitational theory.

%%%%%%%%%%%%%%%%%%%%%%%%%%%%%%%%%%%%%%%%%%%%%%%%%%%%%%%%%
%APPLICATIONS, 4th paragraph. Words: 123

Presently, we are interested in any consistent theory of real Abelian spin-one fields propagating over four-dimensional anti-de Sitter spacetime $AdS_4$ and supporting a black hole.
The reason is that such theories can provide a holographically dual description of a superconductor in $(1+2)$ dimensions, as shown in~\cite{Hartnoll:2008vx}.
This is a pertinent duality, since several unconventional superconductors are layered and much of their physics is three-dimensional
---cuprates and organics, for instance.
The same holds true for the much better understood thin-film superconductors.
Working in the probe limit, where fluctuations of the background and backreaction effects are ignored,
toy models realizing the duality of~\cite{Hartnoll:2008vx} for doped superconductors were proposed in~\cite{Hashimoto:2012pb}.
We point out that these models are multi-field subcases of the Maxwell-Proca theory, requiring $N,M\geq1$.

%%%%%%%%%%%%%%%%%%%%%%%%%%%%%%%%%%%%%%%%%%%%%%%%%%%%%%%%%
%APPLICATIONS, 5th paragraph. Words: 161

    For concreteness, consider the following example, directly drawn from~\cite{Hashimoto:2012pb}: a Maxwell field $A_\mu$ and a Proca field $B_\mu$ with interactions
    \begin{align}
    \mathcal{L}_{\textrm{ex}1}=B^{2}(m^2/2+c\epsilon^{\mu\nu\rho\sigma}A_{\mu\nu}B_{\rho\sigma}),
    \label{Lex1}
    \end{align}
    where $m$ is the mass of the Proca field and $c$ is the coupling constant specifying the interaction strength between the Maxwell and Proca fields.
    Here, indices are raised/lowered with an asymptotically $AdS_4$ black hole metric.
    Notice that, in the flat spacetime limit, (\ref{Lex1}) is contained in (\ref{L0gh}).
    The corresponding action represents a doped superconductor, under the duality relations
    \begin{align}
    A_\mu\longleftrightarrow \overline{\psi}\gamma^{\widetilde{\mu}}\psi, \qquad B_\mu\longleftrightarrow \overline{\psi}_{\textrm{im}}\gamma^{\widetilde{\mu}}\gamma^5\psi_{\textrm{im}}.
    \end{align}
    Namely, the Maxwell field is dual to the usual conducting fermions $\psi$,
    while the Proca field is dual to the (spin-type) impurity fermions $\psi_{\textrm{im}}$.
    Here, $\gamma^{\widetilde{\mu}}$ are the Gamma matrices that generate the Clifford algebra in the three-dimensional spacetime supporting the superconductor,
    and so $\tilde{\mu}=0,1,2$.
    As usual, $\overline{\psi}_{(\textrm{im})}={\psi}_{(\textrm{im})}^\dagger\gamma^0$ and $\gamma^5:=i\gamma^0\gamma^1\gamma^2$.

%%%%%%%%%%%%%%%%%%%%%%%%%%%%%%%%%%%%%%%%%%%%%%%%%%%%%%%%%
%APPLICATIONS, 6th paragraph. Words: 107

We put forward the idea that an extension of the models in~\cite{Hashimoto:2012pb} to the full Maxwell-Proca theory
(or to any $M\geq2$ subcase of it, where our main result (\ref{secconst}) applies) is worthy of attention.
This is because the corresponding action allows for the holographic modeling of a superconductor in the presence of multiple impurities,
all of which are distributed over the entire sample and interact with each other.
It would be interesting to investigate the competition/enhancement effects produced by the interaction of the various impurities
for observables such as the spin susceptibility.
This seems feasible, as enhancing effects in a closely related setup have already been reported~\cite{Musso:2013rnr}.

%%%%%%%%%%%%%%%%%%%%%%%%%%%%%%%%%%%%%%%%%%%%%%%%%%%%%%%%%
%APPLICATIONS, 7th paragraph. Words: 85

{\it Black hole physics.}
Black holes are an innate prediction of General Relativity.
As such, they are gravitational entities, regions of spacetime.
However, when quantum effects are taken into account, black holes are ascribed temperature and entropy, thereby resembling thermodynamical entities.
Attempts to find agreement between the gravitational and quantum field theoretical descriptions of the rich phenomenology of black holes invariably lead to conflict.
Therefore, black holes are intrinsically interesting objects to study, often used as a guide to formulate a theory of quantum gravity.

%%%%%%%%%%%%%%%%%%%%%%%%%%%%%%%%%%%%%%%%%%%%%%%%%%%%%%%%%
%APPLICATIONS, 8th paragraph. Words: 113

The appearance of an explicit mass term $m^2B^{2}/2$ for a Proca field on a dynamical background $g_{\mu\nu}$
hugely limits the existence of regular black hole solutions, as proven in~\cite{Beck}.
However, we already noted that a Proca field can be realized exclusively through derivative self-interaction terms.
This leaves open a possibility~\cite{footnogo} to elude the no-goes of~\cite{Beck}, which was exploited for instance in~\cite{Chagoya:2016aar}.
Moreover, a multi-field scenario is capable of generating hairy solutions.
Consider the case of the Kerr black hole with a complex Proca hair in~\cite{Herdeiro:2016tmi}.
This is a non-interacting $M=2$ subcase of our theory.
The arguments in~\cite{Herdeiro:2016tmi} suggest that our proposed $\mathcal{L}^{(BB)}$ sector is a fertile ground for avoiding the no-goes of~\cite{Beck}.

%%%%%%%%%%%%%%%%%%%%%%%%%%%%%%%%%%%%%%%%%%%%%%%%%%%%%%%%%
%APPLICATIONS, 9th paragraph. Words: 236

Specifically,~\cite{Chagoya:2016aar} focuses on the theory
\begin{align}
\label{Lex2}
\mathcal{L}_{\textrm{ex2}}=\sqrt{-g}\big(M_{\textrm{Pl}}^2R/2-\Lambda-B_{\mu\nu}B^{\mu\nu}/4+\mathcal{L}_{\textrm{cov}}\big),
\end{align}
with $g:=\textrm{det}(g_{\mu\nu})$, $R$ the Ricci scalar, $M_{\textrm{Pl}}$ the Planck mass, $\Lambda$ the cosmological constant and $\mathcal{L}_{\textrm{cov}}$ given by
\begin{align}
\mathcal{L}_{\textrm{cov}}=d_1\Big[\big(\nabla\cdot B\big)^2-\nabla_\mu B_\nu\nabla^\nu B^\mu-B^2R/2\Big].
\end{align}
Here, $d_1\in\mathbb{R}$ is a coupling constant and $\nabla_\mu$ denotes the covariant derivative.
Observe that $\mathcal{L}_{\textrm{cov}}$ is a covariantization of certain interactions $\mathcal{L}_{(2)}^{(BB)}$ in (\ref{sumn}).
In the decoupling limit $B_\mu\rightarrow\nabla_\mu\pi$, this Lagrangian is a particular example of a beyond-Horndeski theory~\cite{Gleyzes:2014dya}
and therefore admits hairy black holes similar to those in~\cite{Babichev:2017guv}.
We propose to supplement the above action with terms inspired by the Maxwell-Proca theory.
Although ghost-freedom of the following proposal remains unchecked, an appealing addition to the above Lagrangian would be
\begin{align}
\label{extraint}
\sqrt{-g}\big(-A_{\mu\nu}A^{\mu\nu}/4+d_2\epsilon^{\mu_1\dots\mu_4}B_{\mu_1}B^\rho\nabla_{\mu_2}B_\rho\nabla_{\mu_3}A_{\mu_4}\big),
\end{align}
with $d_2\in\mathbb{R}$ another coupling constant.
It accounts for the minimal covariant version of a term in $\mathcal{L}_{(2)}^{(AB)}$.
The appeal resides in both the $U(1)$-invariance introduced by the Maxwell field $A_\mu$ and the Maxwell-Proca interaction.
The theory resulting from adding (\ref{Lex2}) and (\ref{extraint}) again evades the axioms of~\cite{Beck} ---due to the  presence of additional global charges--- and is likely to admit new so-called charged-Proca configurations.
In this case, both charged-Proca stars as reported in~\cite{Garcia:2016ldc} and charged-Galileon black holes~\cite{Babichev:2015rva}
would be natural limits of such unprecedented spacetimes.

%%%%%%%%%%%%%%%%%%%%%%%%%%%%%%%%%%%%%%%%%%%%%%%%%%%%%%%%%
%FINAL REMARKS. total words: 272

\section{Final remarks}
\label{finsect}

%%%%%%%%%%%%%%%%%%%%%%%%%%%%%%%%%%%%%%%%%%%%%%%%%%%%%%%%%
%FINAL REMARKS, 1st paragraph. Words: 75

The Maxwell-Proca theory is a non-trivial generalization to multiple fields of the single Proca theory first proposed in~\cite{Tasinato:2014eka}.
It is the result of the formalization of the notion of a Proca field and its constraint algebra underlying~\cite{Tasinato:2014eka}.
The implementation of the (previously unexplored) secondary level of the algebra on a Lagrangian yields the differential relations in (\ref{secconst}).
Any multi-Proca theory must satisfy these relations to avoid ghosts.
Then, the algebra closes automatically at tertiary level.

%%%%%%%%%%%%%%%%%%%%%%%%%%%%%%%%%%%%%%%%%%%%%%%%%%%%%%%%%
%FINAL REMARKS, 2nd paragraph. Words: 50

The completeness claim relies on the following axioms:
we work on Minkowski spacetime, we consider first-order Lagrangians and we focus on real Abelian spin-one fields.
While it is possible to circumvent our axioms and generate beyond-Maxwell-Proca theories, this is not straightforward.
We comment on the extensions of our axiomatization elsewhere~\cite{PRD}.

%%%%%%%%%%%%%%%%%%%%%%%%%%%%%%%%%%%%%%%%%%%%%%%%%%%%%%%%%
%FINAL REMARKS, 3rd paragraph. Words: 74

The given two applications in no way exhaust the potential of the Maxwell-Proca theory.
Other implications can also be entertained, for instance in the context of cosmology, hydrodynamics and non-linear optics.
Observe that the above described (non-trivial) coupling to gravity is required for applications in gravity and cosmology.
In holographic condensed matter, this coupling is used when working beyond the probe limit.
For all these disciplines, the Maxwell-Proca theory constitutes an essential first step.

%%%%%%%%%%%%%%%%%%%%%%%%%%%%%%%%%%%%%%%%%%%%%%%%%%%%%%%%%
%FINAL REMARKS, 4th paragraph. Words: 71

In the proposed applications, the multi-field feature is cornerstone.
At least one Maxwell and one Proca are required to holographically model a doped (unconventional) superconductor.
If various Procas are considered, enhancement/competition effects of dopants can be thus studied.
In the black holes case, the Maxwell-Proca theory encloses at least three proven ways to evade no-hair theorems.
Further, its vast interaction terms should constitute a fruitful source of novel hairy solutions.

\vspace{1\baselineskip}

The authors thank Axel Kleinschmidt for a fruitful discussion, which gave us the idea for this work.
We are grateful to Chrysoula Markou for enlightening conversations in the early stages of
the project. VED thanks Nicol\'{a}s Coca L\'{o}pez for lucid explanations regarding superconductors. JAMZ
is thankful to  E. A. Ay\'{o}n-Beato and D. F. Higuita-Borja for the computing power granted at ZymboLab (CINVESTAV).
The work of JAMZ is partially supported by the ``Convocatoria Max-Planck-CONACyT 2017 para estancias postdoctorales'' fellowship.
This work is supported by a grant from the Max Planck Society.

%%%%%%%%%%%%%%%%%%%%%%%%%%%%%%%%%%%%%%%%%%%%%%%%%%%%%%%%%
%FOOTNOTES. Words: 155

%%%%%%%%%%%%%%%%%%%%%%%%%%%%%%%%%%%%%%%%%%%%%%%%%%%%%%%%%
%TOTAL. Words: 864+1269+1172+256+155=3732 (Must be under 3750 )


\begin{thebibliography}{99}
\bibitem{footLAB1}
This is because any interaction term involving a Maxwell field $A_\mu$ contains at least one derivative
within the corresponding field strength $A_{\mu\nu}$.
For the term to belong in $\mathcal{L}_{(1)}$, this derivative must be the only one present.
But in that case, the term belongs in $\mathcal{L}_{(0)}$ by definition.
\end{thebibliography}
\end{document}